**Clinical test cases for commissioning, QA, and benchmarking of model-based dose calculation algorithms in $^{192}$Ir HDR gynecologic tandem and ring brachytherapy**

V. Peppa (1 and 2), M. Robitaille (3), F. Akbari (4), S. A. Enger (3), R. M. Thomson (4), F. Mourtada (5), G. P. Fonseca (6)

[1]Medical Physics Laboratory, Medical School, National and Kapodistrian University of Athens, Athens, Greece

[2]Radiotherapy Department, General Hospital of Athens Alexandra, Athens, Greece

[3]Medical Physics Unit, Department of Oncology, Faculty of Medicine, McGill University, Montréal, Québec, Canada

[4]Carleton Laboratory for Radiotherapy Physics, Physics Department, Carleton University, Ottawa, Ontario, Canada

[5]Department of Radiation Oncology, Sidney Kimmel Cancer Center, Thomas Jefferson University, Philadelphia, Pennsylvania, USA

[6]Department of Radiation Oncology (MAASTRO), GROW School for Oncology and Developmental Biology, Maastricht University Medical Centre+, Maastricht, Netherlands

## ABSTRACT

**Purpose:** To develop clinically relevant test cases for commissioning Model-Based Dose Calculation Algorithms (MBDCAs) for $^{192}$Ir High Dose Rate (HDR) gynecologic brachytherapy following the workflow proposed by the TG-186 report and the WGDCAB report 372.

**Acquisition and Validation Methods:** Two cervical cancer intracavitary HDR brachytherapy models were developed based on a real patient, using either uniformly structured regions or realistic segmentation. The patient's computed tomography (CT) images were processed, converted to a series of Digital Imaging and Communications in Medicine (DICOM) CT images, and imported into two Treatment Planning Systems (TPSs), the Oncentra Brachy and BrachyVision. The original segmentation of the clinical case was augmented to enable a thorough dosimetric analysis. The actual clinical treatment plan was generally maintained, with the source replaced by a generic $^{192}$Ir HDR source. Dose to medium in medium calculations were performed using the MBDCA option of each TPS, and three different Monte Carlo (MC) simulation codes. MC results demonstrated agreement within statistical uncertainty, while comparisons between the commercial TPS MBDCAs and a general-purpose MC code highlighted both the advantages and limitations of the studied MBDCAs, suggesting potential approaches to overcome the challenges.

**Data Format and Usage Notes:** The datasets for the developed cases are available online at https://doi.org/10.5281/zenodo.15720996. The DICOM files include the treatment plan for each case, TPS, and the corresponding reference MC dose data. The package also contains a TPS- and case-specific user guide for commissioning the MBDCAs, as well as files necessary to replicate the MC simulations.

**Potential Applications:** The provided datasets and proposed methodology can serve as a commissioning framework for TPSs that employ MBDCAs, as well as a benchmark for brachytherapy researchers using MC methods and MBDCA developers. They also facilitate intercomparisons of MBDCA performance and provide a quality assurance resource for evaluating future TPS software updates.

## 1 INTRODUCTION

Commissioning, as well as systematic monitoring and evaluation, are integral parts of a phased execution process to successfully implement innovation in radiotherapy.[1] Model-based dose calculation algorithms (MBDCAs) clearly marked an innovation in brachytherapy, paving the way for improved absorbed dose predictions that support individualization of treatments, safe adoption of new sources, applicators and techniques, establishment of robust dose–response relationships, and enhanced dose reporting accuracy, as current protocols are based on TG-43 formalism, which relies on the superposition of single source dosimetry data pre-calculated in an unbounded water geometry, thereby neglecting factors such as tissue heterogeneities, finite patient dimensions, and the presence of applicators.[2–4]

MBDCAs can calculate dose in computational models defined by patient 3D imaging, hence accounting for tissue and applicator heterogeneities. Besides the "gold standard" of Monte Carlo (MC) simulation, examples include the two algorithms that have been commercially available for clinical use in $^{192}$Ir High Dose Rate (HDR) brachytherapy applications for over a decade,[5] namely Acuros BV™ (Varian Medical Systems, Palo Alto,CA) and TG-186 ACE™ (Elekta Brachy, Veenendaal, The Netherlands). Acuros BV and ACE are based on a grid-based linear Boltzmann transport equation solver[6,7] and the collapsed cone superposition method,[8] respectively. Their implementation in treatment planning systems (TPSs) was initially validated through independent comparisons of the commercial MBDCAs to MC simulation or experimental results,[5,9–12] and corresponding studies continue to date for different clinical sites.[13–19] It was readily realized however that such validation studies are not feasible as commissioning and periodic quality assurance tools in the clinical setting.[20] Independent research efforts setting forth end-user-oriented data and procedures[21] were superseded by the concerted effort of the working group on dose calculation algorithms in brachytherapy (WGDCAB) to facilitate and standardize the commissioning workflow proposed by the joint European Society for RadioTherapy and Oncology (ESTRO), Australasian Brachytherapy Group (ABG), American Brachytherapy Society (ABS), and American Association of Physicists in Medicine (AAPM) Task Group 186 (TG-186) report.[20]

The WGDCAB has developed test case datasets available through the Brachytherapy Source Registry[22–26] as well as a report on the detailed practical implementation of the commissioning process, including quantitative goals,[2] and has recently submitted a study on the development of 3D reference dosimetric datasets for permanent implant prostate brachytherapy.

Besides the continuing importance of brachytherapy for both definitive and adjuvant treatment of cervical, endometrial, and vaginal cancers, patient equivalent test case data for level 2 commissioning[20] in $^{192}$Ir brachytherapy to date are limited to interstitial HDR breast brachytherapy.[24]

This work presents the first clinically oriented test case datasets for $^{192}$Ir intracavitary brachytherapy for gynecologic cancers. It aims to provide a vendor-neutral reference dataset for commissioning and validation, serving as a foundation for benchmarking TPS algorithms and comparing dosimetric models, while supporting users in implementing approaches beyond TG-43, such as tissue classification and CT-based material and density mapping. The clinical relevance and magnitude of differences between TG-43 and MBDCA approach have been reported for multiple patients accounting for applicators, contracts agents, patient anatomy and composition [27–29] and are out of the scope of this dataset article.[30]

The current study represents an independent investigation by members of the joint AAPM/ESTRO/ABS/ABG WGDCAB and not a societal recommendation.

# 2 ACQUISITION AND VALIDATION METHODS

## 2.1 Patient computational models

Two models were generated in this work based on an anonymized cervical cancer case, representative in terms of age, weight, bladder and rectum volume, treated with intracavitary $^{192}$Ir HDR brachytherapy using a tandem and ring applicator.

The first model (test case A) is based on Regions of Interest (ROIs) of uniform density, an approach previously used in the literature to facilitate the interpretation of MBDCA commissioning results.[2] A commercial software package (MATLAB 2020a, The MathWorks Inc., Natick, MA) was used for processing of the Computed Tomography (CT) images and the structure set of the clinical case. Specifically, a nominal Hounsfield Unit (HU) was assigned to each delineated structure including the target, bladder, rectum, bowel and sigmoid. This was equal to the mean HU value of the voxels encompassed by the contour of the given ROI. It should be noted that the contrast agent partially filling the rectum and bladder of the original case was excluded from the calculation of the mean HU value and subsequently overridden. In order to generate ROIs corresponding to bony structures, a HU threshold appropriate to exclude bone marrow and cartilage was applied, obtaining a skeletal tissue with a medium density (1.4049 g/cm$^3$) within the broad spectrum of densities (from 0.98 g/cm$^3$ for yellow marrow to 1.92 g/cm$^3$ for cortical bone) and material compositions found in the literature for bones.[31] It should be mentioned that, mass density to HU conversion during model preparation was performed for each ROI using the default CT calibration curve of BrachyVision TPS. Given the limited material composition schemes included in the commercially available TPSs for densities in the range of bones (cortical bone in Oncentra Brachy,[20] cartilage and cortical bone in BrachyVision[32]) a skeletal tissue with elemental properties differing from cortical bone was intentionally incorporated to highlight potentially increased dosimetric differences arising from unrealistic material composition assignments,[33] particularly due to variations in calcium content.[34] The tandem and ring applicator was contoured in Oncentra Brachy v4.6 and processed so that its HU corresponds to the nominal density of polyphenylsulfone (PPSU), which is included in the material libraries of both Oncentra Brachy and BrachyVision. Specifically, a mass density of 1.2901 g/cm$^3$ was considered, corresponding to the nominal PPSU mass density value defined in the Oncentra Brachy TPS, while the source path within the applicator was modeled as water to account for the presence of plastic catheters. For simplicity, air was not included in the source path, an approximation that is nevertheless not expected to introduce bias in the dosimetric comparisons presented in this work. Table 1 summarizes the resultant HUs applied to each ROI and the associated mass density values based on the CT calibration curve of the BrachyVision TPS. Figure 1 shows an axial image of the model.

Table 1. HUs and corresponding density values assigned to the ROIs generated for test case A.

| ROI | HU | Density (g/cm$^3$) |
|---|---|---|
| Target | 40 | 1.0417 |
| Bladder | 93 | 1.0781 |
| Rectum | 37 | 1.0385 |
| Bowel | -71 | 0.9630 |
| Sigmoid | -48 | 0.9750 |
| Bones | 646 | 1.4049 |
| Tandem and ring applicator | 453 | 1.2901 |
| External (body) | 0 | 1.0000 |
| Air | -992 | 0.0012 |

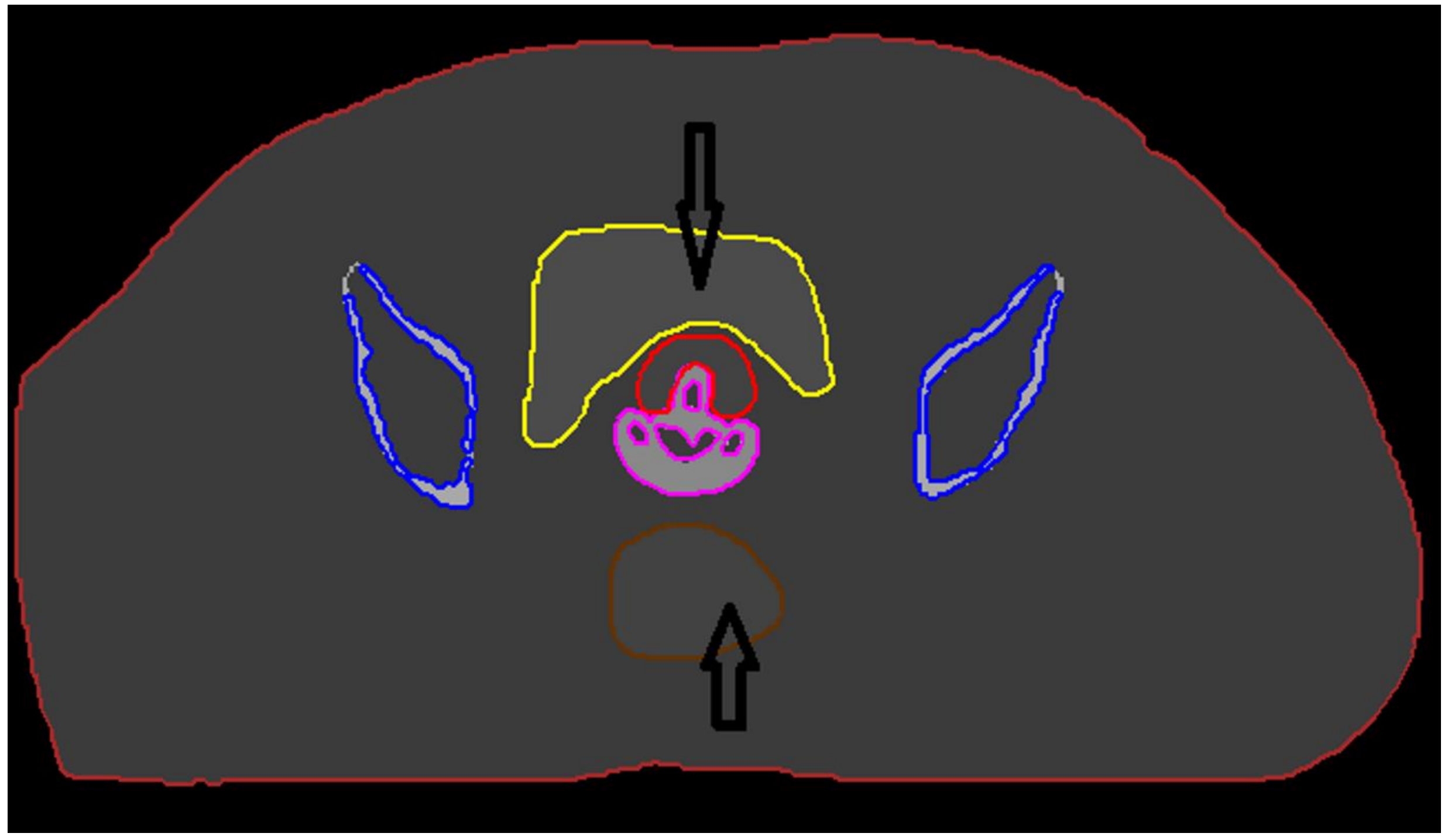

Figure 1. An axial image of the patient phantom model generated for test case A depicting the target (red contour), bladder (yellow contour), rectum (brown contour), bones (blue contour) and applicator (magenta contour). Black arrows indicate regions within the bladder and rectum that initially contained contrast agent.

The second model (test case B) adopts a more realistic approach to validate the capability of MBDCAs to accurately account for heterogeneities across the full range of HUs encountered in actual clinical scenarios. Test case B differs from A in that the HUs of the original case were preserved in the axial CT images processed using MATLAB, except for voxels in the bladder, the applicator, and the rectum. The bladder and the applicator were handled as in test case A. For the rectum a more realistic scenario was adopted wherein the air voxels occupying the patient's rectum were incorporated into the existing uniform structure generated for test case A, resulting in a ROI partially occupied by air. Figure 2 presents an axial image of the model generated for test case B.

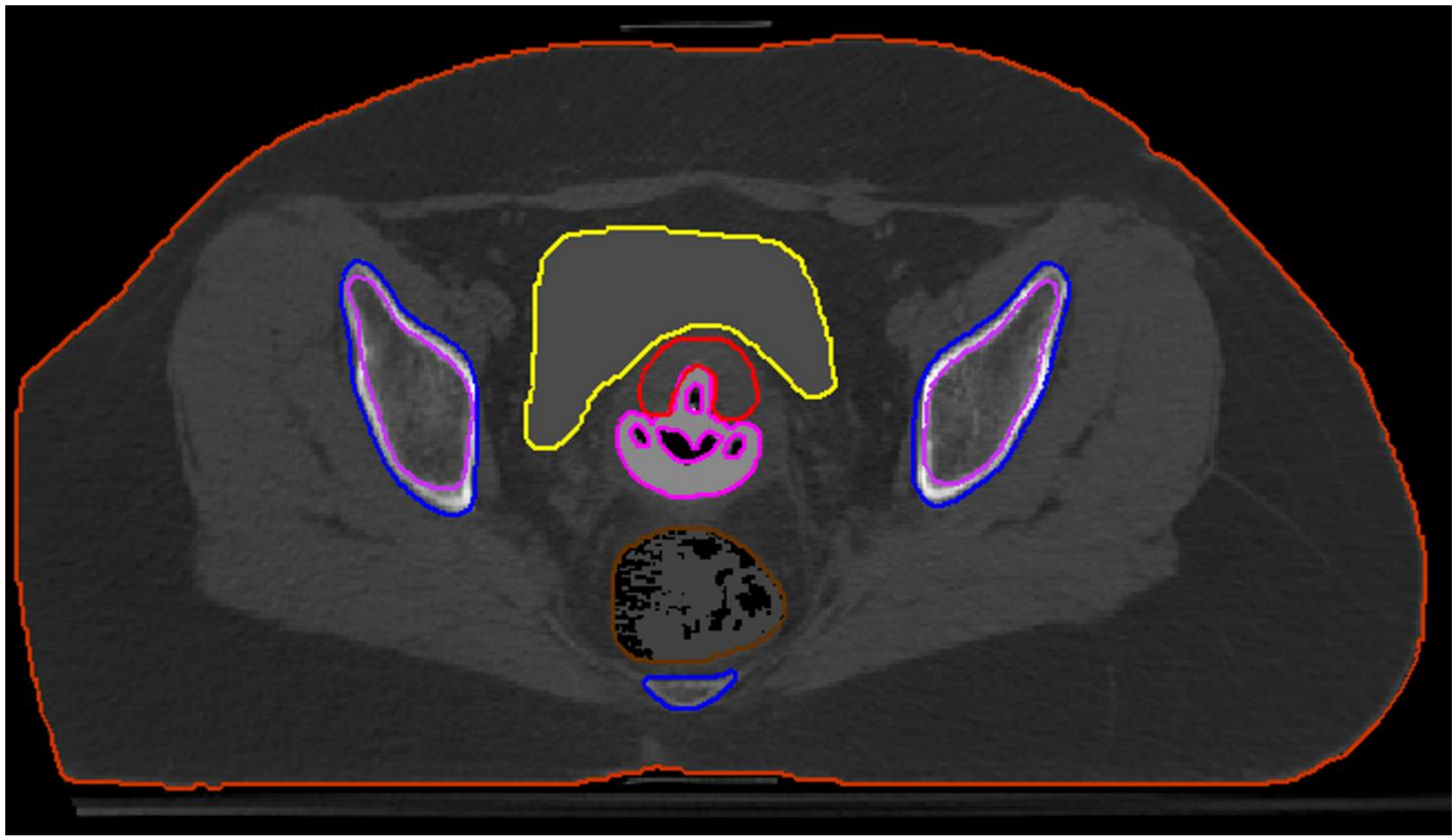

Figure 2. Axial image of the patient phantom model generated for test case B depicting the target (red contour), bladder (yellow contour), rectum (brown contour), pelvic bones (blue contour), marrow (purple contour) and applicator (magenta contour).

After their configuration, both patient models were written to a series of DICOM CT images using MATLAB. The resolution was identical to that of the original patient imaging (512 x 512 x 111, with voxel dimensions of 0.935 × 0.935 × 3 $mm^3$), with no change made to the existing field values of the original anonymized dataset.

## 2.2 Treatment planning

### 2.2.1 Oncentra Brachy TPS

The CT images of the computational models prepared for test cases A and B, along with the RT structure set and RT plan file of the clinical case, were imported into Oncentra Brachy v4.6. Additional contouring was performed for both test cases to enhance the performance of the Advanced Collapsed Cone Engine (ACE) in Oncentra Brachy (wherein elemental composition assignment is contour-based), and to enable a thorough dosimetric analysis that could provide deeper insight into how tissue elemental composition affects MBDCA performance. For test case A, the contour of the bones was added to the existing structure set using the automatic segmentation tools of the TPS. For test case B, an artificial intelligence-powered software tool (ART-Plan™, TheraPanacea) was employed for automatic contour delineation of bone anatomical regions. This resulted in four ROIs (bone marrow, bilateral femoral heads, and pelvic bones including the two hip bones, each composed of the ilium, ischium, and pubis, the sacrum, and the coccyx) which were added to the structure set of test case B. For both test cases, mass density was assigned to each structure using the HU-based option of the TPS, whereby individual voxel density was derived from CT data and a user-defined calibration, which was set to the default one in BrachyVision for consistency purposes between the two TPSs. Material assignment was performed by the user defined option based on elemental compositions from ICRU 46.[35] Table 2 summarizes the material assignment used in Oncentra Brachy for the structures in test cases A and B.

Table 2. Material assignment to the delineated structures of test cases A and B in Oncentra Brachy TPS, with their corresponding elemental tissue compositions.

| | Elemental Composition | | % mass[20] | | | | |
|---|---|---|---|---|---|---|---|
| ROI | Test case A | Test case B | H | C | N | O | Z>8 |
| Target<br>Rectum<br>Bowel<br>Sigmoid | Female soft tissue | Female soft tissue | 10.6 | 31.5 | 2.4 | 54.7 | Na(0.1), P(0.2), S(0.2), Cl(0.1), K(0.2) |
| Bones<br>Pelvic Bones | Cortical bone<br>- | -<br>Cortical bone | 3.4 | 15.5 | 4.2 | 43.5 | Na(0.1), Mg(0.2), P(10.3), S(0.3), Ca(22.5) |
| Femoral Heads<br>Bone Marrow | -<br>- | Mean gland | 10.6 | 33.2 | 3.0 | 52.7 | Na(0.1), P(0.1), S(0.2), Cl(0.1) |
| Applicator | PPSU | PPSU | 4.0 | 72.0 | | 16.0 | S(8.0) |
| Bladder<br>External (body) | Water | Water | 11.2 | | | 88.8 | |

For treatment planning, the source dwell positions and planning aim (5 Gy per fraction to ICRU point A[36]) of the original plan were maintained, with relative weights set to unity for simplicity. The tandem and ring applicator used for the clinical case was represented by its corresponding structure (i.e. an applicator library model was not used). The source in the original treatment plan was switched to the WG generic $^{192}$Ir HDR source,[37] resulting in a negligible change of less than 0.01% in the Total Reference Air Kerma (TRAK). Dose to water in water ($D_{w,w}$) and dose to medium in medium ($D_{m,m}$) were calculated with an isotropic dose grid resolution of 1 mm using the TG-43-based algorithm and the ACE, respectively. Both the standard accuracy (SA) and high accuracy (HA) options were employed for the latter. These options define the number of transport directions used for the first and multiple scatter dose calculations, which were 320 and 180 for the SA level and 720 and 240 for the HA level, respectively, based on the number of dwell positions in the treatment plan. Additionally, the inherent voxel size used for dose calculation with ACE depends on the selected accuracy level and increases with the margin of the bounding boxes surrounding the dwell positions, with voxel sizes of 1, 2, 5 and 10 mm for the SA level and 1, 2, and 5 mm for the HA level. The corresponding bounding box margins are 1, 8, 20 and 50 cm for the SA option and 8, 20 and 50 cm for the HA option. The resulting calculation times for the test cases A and B are presented in Table 3.

### 2.2.2 BrachyVision TPS

The CT images and RT structure sets exported from Oncentra Brachy for each test case were then imported into BrachyVision v.16.1. The import of a structure set generated in one TPS to another had a minor effect in terms of structure coordinates and volumes. The latter were evaluated independently using a MATLAB function (*inpolygon)* to verify the number of voxels within each contour, showing agreement within 1% in the exports from the two TPSs. Treatment planning in BrachyVision for test cases A and B was performed similarly to that in Oncentra Brachy, using the WG generic $^{192}$Ir HDR source with a planning aim of 5 Gy per fraction to ICRU point A.[36] It should be noted that although the contours of the structures were automatically imported into BrachyVision TPS, treatment plan parameters such as dwell positions, catheter points and dwell times were not parsed automatically. Despite the technical differences between the two TPSs, the effort made to manually replicate the Oncentra Brachy-based treatment plan in terms of dwell positions, direction cosines and TRAK resulted in differences of less than 0.1 mm, 0.07 and 0.1%, respectively, thus ensuring uniformity between the treatment plans in the two TPSs. Given that TPSs do not provide information on source orientation, agreement in direction cosines was independently verified using the coordinates of the two reconstructed catheter points immediately before and after each dwell position along the catheter path.[38] Consistent with the Oncentra Brachy plan, the 25 dwell positions were equally weighted, and the applicator was represented by its corresponding structure.

The dose reporting grid resolution for $D_{w,w}$ and $D_{m,m}$ calculations using the TG-43-based algorithm and Acuros BV, respectively, was also the same (1 mm isotropic), yielding a calculation time of approximately 4 min for the latter (Table 3). Material assignment in Acuros BV is automated based on a density lookup table with tissue elemental compositions taken from ICRP 23.[32] The difference of MBDCA results from the two TPSs due to differences in tissue elemental compositions is however expected to be small for $^{192}$Ir dosimetry.[24] It should be noted that dose calculations using Acuros BV are performed with the pre-defined material density of PPSU, which is set to 1.30 g/cm$^3$.

The CT images, RT structure set, RT plan and RT dose files exported in *.DCM, *RS, *RP and *RD format comprise TPS- and test case- specific data of the datasets.

Table 3. Time required for MBDCA calculations for test cases A and B using Oncentra Brachy and BrachyVision TPSs.

| Calculation time | | |
|---|---|---|
| OncentraBrachy TPS | | BrachyVision TPS |
| ACE (HA) | ACE (SA) | Acuros BV |
| ~ 1 h | ~ 5 min | ~ 4 min |

## 2.3 Monte Carlo simulations

In order to attain “reference” status for the MC dose distributions included in the datasets, [2,24,25] independent simulations were performed for both test cases using three different codes, MCNP v.6.2,[39] RapidBrachyMCTPS (Geant4),[40,41] and egs_brachy,[42] for validation purposes. The MCNP input file was prepared using BrachyGuide[38] to parse treatment plan information exported in DICOM RT format. BrachyGuide is freely available for download at https://mpl-en.med.uoa.gr/downloads/. egs_brachy calculations were carried out using eb_gui,[43] a free and open-source software tool developed to facilitate fast Monte Carlo simulations of brachytherapy treatment plans based on egs_brachy, a code designed for dose calculations in brachytherapy applications. The eb_gui software can be accessed at https://github.com/clrp-code/eb_gui. Details of the MC simulations performed are presented in Table 4 following the RECORDS guidelines (improved Reporting of montE CarlO RaDiation transport Studies).[44]

In view of the negligible differences in the dwell positions, direction cosines and TRAK values between the two TPSs, a single MC input file was generated per test case for each code using the average values. This approach, aimed at reducing the amount of data to be distributed,

Table 4. Summary of methods used for Monte Carlo simulations of this work following the TG-268 template [44].

| **Code, version** | MCNP6 v.6.2[39]<br>BrachyGuide v.1.0[38] | Geant4 11.2[45,46]<br>RapidBrachyMCTPS[40,41] | egs_brachy [42,47]<br>eb_gui[43] |
|---|---|---|---|
| Validation | [14,17,19,21,38,48,49] | [40] | [42,43,50] |
| Timing | 2.8 days per case using a local server equipped with two 6-core CPUs (24 computational threads) clocked at 2.3 GHz | 20 hours per case on 2 x Intel Gold 6148 Skylake at 2.4 GHz | 10 days per case on AMD Ryzen 9 5800X. 4.20 GHZ CPU |
| Source description | WG generic $^{192}$Ir source [37] represented by a phase space file of 8 x $10^7$ initially emitted photons emerging from the source [51] | WG generic $^{192}$Ir source [37] | WG generic $^{192}$Ir source [37] |
| Cross-sections | EPDL97 [52] | EPDL97 [52] | XCOM [53] |
| Transport parameters | Photons transported to 1 keV; no electron transport | Production cut: max (1mm, 0.99 keV); no electron transport | Photons transported to 1 keV; no electron transport |
| Variance reduction | - | Track Length estimator using mass-energy absorption [33,54] | - |
| Scored quantities | Absorbed dose approximated by collision kerma calculated using a track length estimator | | |

| # histories/ Statistical uncertainty | $4.8\times10^9$ histories, up to 1.0% within the target, 2.1% within the rectum, 1.7% within the bladder, 2.2% within the sigmoid, 2.4% within the bowel and 4.2% within the bones | $2.0\times10^8$ histories, up to 0.6% within the target, 1.7% within the rectum, 1.2% within the bladder, 2.1% within the sigmoid, 3.0% within the bowel and 9.5% within the bones | $5\times10^9$ histories, up to 0.2% within the target, 0.5% within the rectum, 0.4% within the bladder, 0.6% within the sigmoid, 0.7% within the bowel and 2.5% within the bones |
|---|---|---|---|
| Statistical methods | History-by-history | | |
| Postprocessing | MCNP reference results were interpolated to match the spatial resolution of the RT dose exports (1 $mm^3$ isotropic) and written to *.RD format files compatible with each TPS | | |

appeared to have a marginal dosimetric impact on the corresponding MCNP simulation results, which was limited to the voxels occupied by the source. It should be noted that the source direction was obtained from the coordinates of the two catheter points closest to each dwell position. Mass density was assigned to each scoring voxel ($0.935 \times 0.935 \times 1$ $mm^3$) using the default CT calibration curve of BrachyVision TPS, ensuring consistency with treatment planning and MBDCA calculations. For tissue-like materials, elemental composition in MCNP and RapidBrachyMCTPS was assigned to each voxel based on its mass density, using a look-up table of 23 human composition bins.[31,34] Elemental composition was also assigned on a voxel-by-voxel basis in eb_gui, through a user-defined structured tissue assignment scheme (TAS)[55,56] that incorporates structure contours, voxel mass density and the same look-up table of composition bins. For the voxels representing the tandem and ring applicator, a mass density of 1.2951 $g/cm^3$ was applied, calculated as the average of the nominal PPSU mass density values considered in Oncentra Brachy (1.2901 $g/cm^3$) and BrachyVision TPS (1.3000 $g/cm^3$). The dosimetric impact of the PPSU mass density variation between the two TPSs was assessed through MC simulation with MCNP, revealing deviations of less than 1% in the applicator structure, while remaining within MC

Type A uncertainty elsewhere. Details of the parameters used in this work for the MC simulations are provided in Tables S1, S2 and S3 of the Supplementary Material.

## 2.4 MC data validation

MCNP results are included in the datasets as reference MC dose distributions, as this code was also employed to assess potential dosimetric inaccuracies introduced by the dwell positions, direction cosines, TRAK values, and PPSU mass density compromises necessary to generate a single MC input file per case for both TPSs, without bias against either of the other two codes. The reference results were validated against the corresponding RapidBrachyMCTPS and eb_gui dose distributions.

### 2.4.1 Test case A

Figure 3 presents colormaps of the percentage local differences $\left(\%\Delta D_{LOCAL} = \frac{D_{eval}(x,y,z)-D_{ref}(x,y,z)}{D_{ref}(x,y,z)} \times 100\right)^2$ of RapidBrachyMCTPS and eb_gui relative to the reference dataset for test case A. The corresponding distributions of % $\Delta D_{LOCAL}$ calculated within each structure are also included in the same Figure in the form of box plots, while the median % $\Delta D_{LOCAL}$ values and 95% percentile ranges are presented in Table 5. In Figures 3(a) and 3(c), agreement within Type A uncertainties can be observed between the three MC codes, except for specific voxels within the target that are partially occupied by the source. Differences in these voxels (up to approximately 20% in Figure 3(b) and 5% in Figure 3(d)) are attributed to the different approaches by each code. Specifically, no mass correction was made in MCNP, whereas RapidBrachyMCTPS uses the layered mass geometry technique to model the overlap of the source with the patient model,[57] and eb_gui applies the "volume correction" method, in which the dose is scored solely to the portion of the voxel not occupied by the source.[42] In Figure 3(b), the distributions of % $\Delta D_{LOCAL}$ between RapidBrachyMCTPS and MCNP appear normally distributed with median values in Table 5 ranging from -0.06% for the target to -0.56% for the bones. A similar trend can be observed in Figure 3(d) and Table 5 for eb_gui, where the median % $\Delta D_{LOCAL}$ relative to MCNP range from -0.42% for the target to -0.59% for the bones. For both codes, the range of the differences increases with increasing distance from the implant, reflecting the inherent rise in MC Type A uncertainty (see

Table 4), with the 95% percentile range for RapidBrachyMCTPS being [-0.92%, 0.77%] for the target and [-4.87%, 3.94%] for the bones. The corresponding results for eb_gui were [-0.98%, 0.12%] for the target and [-2.60%, 1.41%] for the bones. These findings align with the comparison of Dose Volume Histogram (DVH) indices in Table S4 of the Supplementary Material calculated from MCNP, RapidBrachyMCTPS and eb_gui dose distributions using MATLAB function *inpolygon* to segment the structures, where the % $\Delta D_{LOCAL}$ agreement between RapidBrachyMCTPS and eb_gui relative to the MCNP results is within 0.68% and 0.57%, respectively.

### 2.4.2 Test Case B

In Figure 4, similar results can be observed for the comparison of RapidBrachyMCTPS and eb_gui dose distributions relative to MCNP for test case B. Agreement between the three codes is again within MC Type A uncertainty (Figures 4(a) and 4(b)) apart from voxels partially occupied by the source where maximum differences within the target (Figures 4(c) and 4(d)) are comparable to those for test case A. Elevated dosimetric differences between the eb_gui and reference results were also observed in a limited number of peripheral bone voxels, due to the contour priority approach used by eb_gui to handle elemental composition, which favored PPSU over skeletal material in voxels of identical mass density, without however compromising the overall agreement between the two codes within the bones. The distributions of % $\Delta D_{LOCAL}$ in each structure shown in Figures 4(c) and 4(d), appear normally distributed with a wider spread as distance from the implant increases. The median values and 95% percentile ranges of % $\Delta D_{LOCAL}$ span from -0.61% [-2.04%, 0.86%] for the target to -0.52% [-4.68%, 3.78%] for the pelvic bones for RapidBrachyMCTPS and from -0.46% [-1.75%, 0.89%] for the target to -0.48% [-3.73%, 3.14%] for the pelvic bones for eb_gui (see Table 5). These results are in accordance with the comparison of DVH indices presented in Table S5 of the Supplementary Material, where an agreement within 1.05% and 1.09% can be seen between RapidBrachyMCTPS and eb_gui results, respectively, relative to MCNP across all the considered structures.

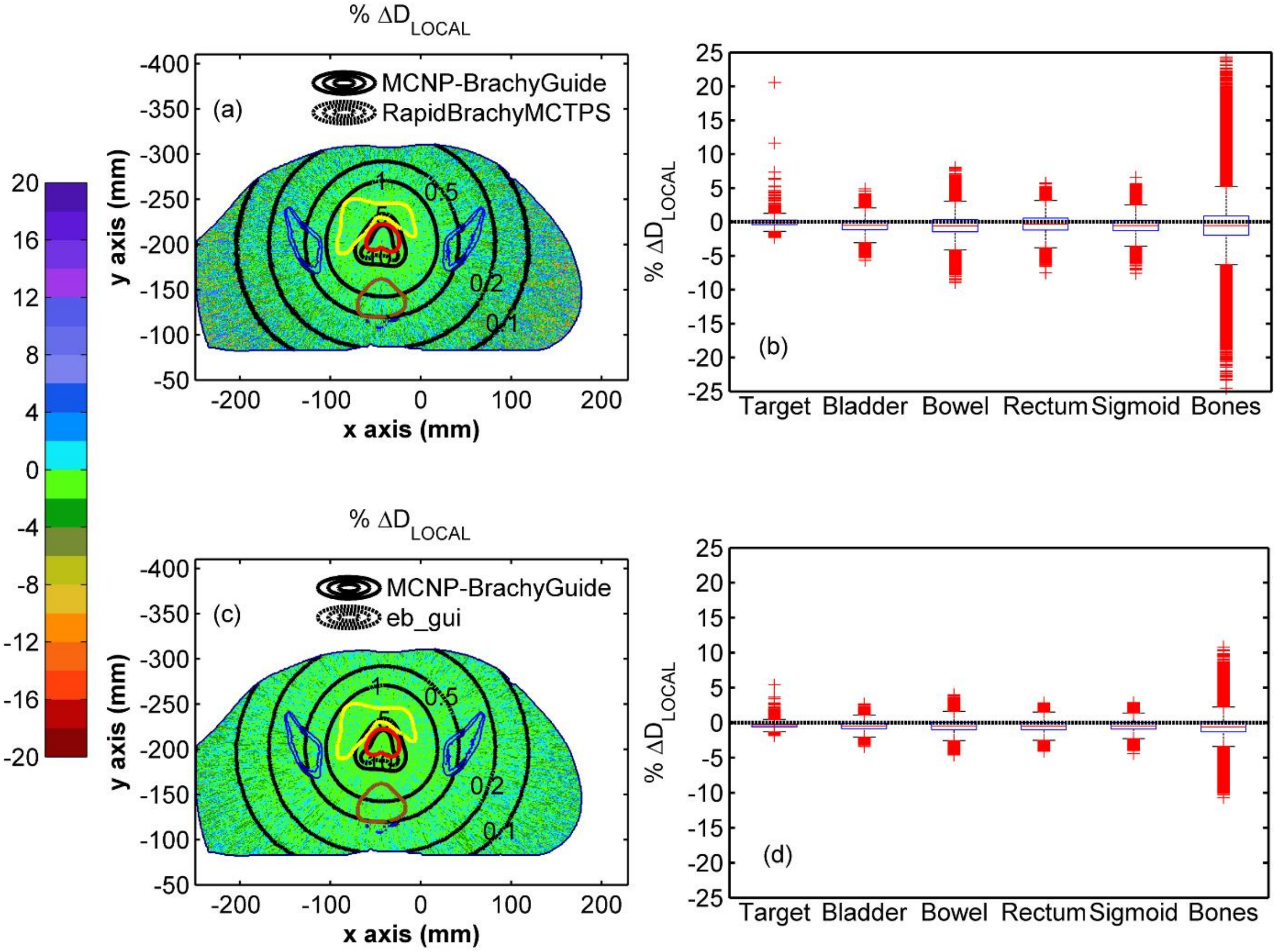

Figure 3. Colormap representations of % $\Delta D_{LOCAL}$ between (a) RapidBrachyMCTPS and (c) eb_gui, and reference MCNP results on an axial slice of test case A, with selected isodose lines in Gy (0.1, 0.2, 0.5, 1.0, 5.0, 10.0) superimposed (red contour: target, yellow contour: bladder, brown contour: rectum, blue contour: bones). Differences in isodoses are not visible due to their overlap. Corresponding box and whiskers plots of % $\Delta D_{LOCAL}$ between (b) RapidBrachyMCTPS or (d) eb_gui, and reference MCNP results are presented for the target, bladder, bowel, rectum, sigmoid, and bones. Whiskers extend to 1.5 times the interquartile range and the red columns shown are formed by the overlapping of outliers.

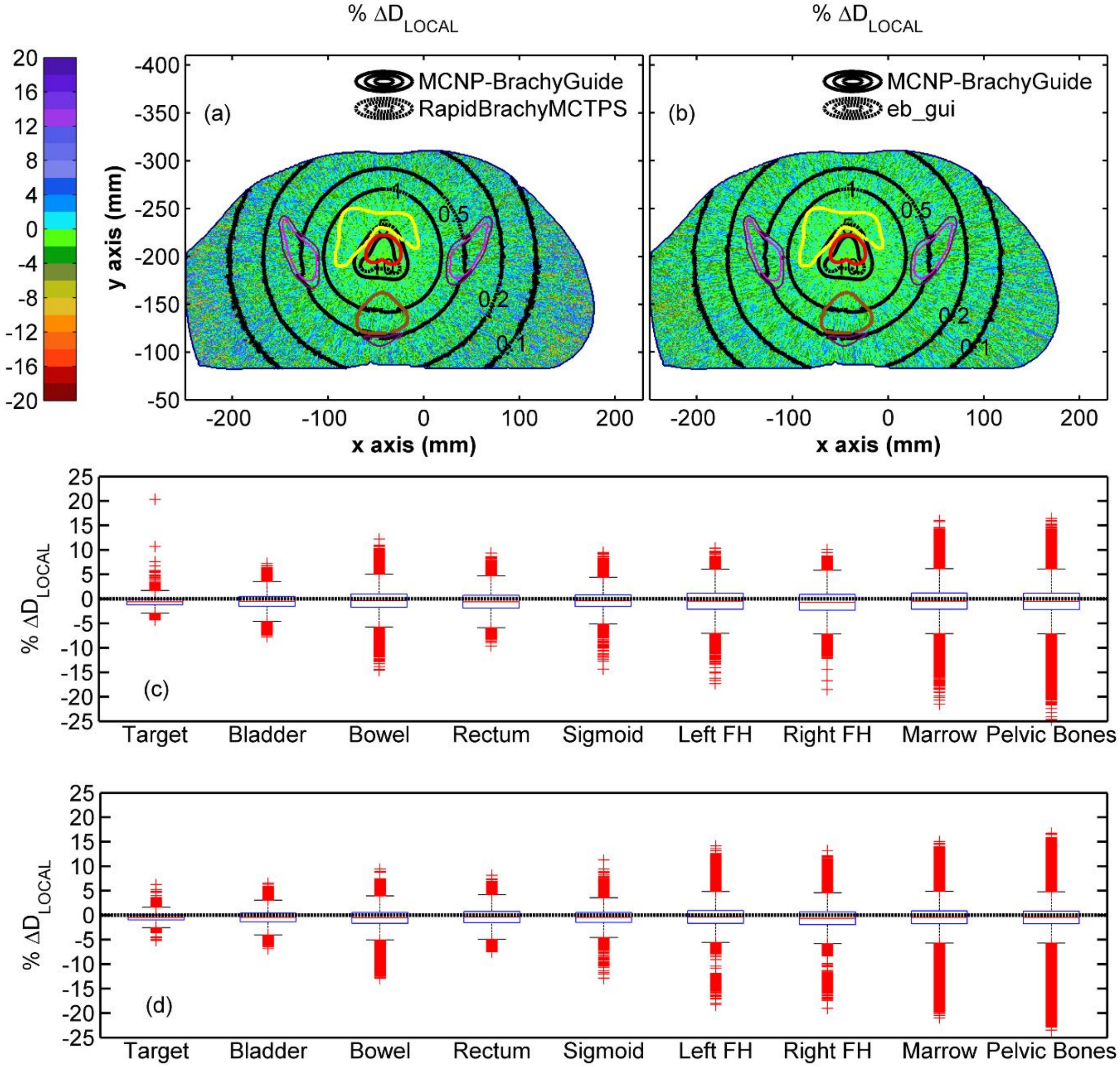

Figure 4. Colormap representations of the % $\Delta D_{LOCAL}$ between (a) RapidBrachyMCTPS and (b) eb_gui, and MCNP results on an axial slice of Test case B, with selected isodose lines in Gy (0.1, 0.2, 0.5, 1.0, 5.0, 10.0) superimposed (red contour: target, yellow contour: bladder, brown contour: rectum, magenta contour: marrow, blue contour: pelvic bones). Differences in isodoses are not visible due to their overlap. Corresponding box and whiskers plots of the % $\Delta D_{LOCAL}$ between (c) RapidBrachyMCTPS and (d) eb_gui with MCNP results are presented for the target, bladder, bowel, rectum, sigmoid, left femoral head, right femoral head, marrow and bones. Whiskers extend to 1.5 times the interquartile range and the red columns shown are formed by the overlapping of outliers.

## 2.5 TPS MBDCA data validation

Reference MBDCA results obtained by ACE (Oncentra Brachy v4.6) and Acuros BV (BrachyVision v16.1) for test cases A and B are also included in the datasets to facilitate phase 1 of the commissioning workflow presented in Figure 2 of the WGDCAB report 372.[2] Their validation against reference MC data is required: (a) as a sanity check, whereby findings should be consistent with initial MBDCA validation studies in the literature, (b) to illustrate results expected from the use of the datasets for commissioning purposes, and (c) to establish quantitative goals for phase 2 of the WGDCAB report 372 commissioning workflow.

### 2.5.1 Test case A

Figure 5 presents the comparisons of ACE (HA) and reference, MCNP dose data for test case A. The isodose lines and % $\Delta D_{LOCAL}$ in Figure 5(a) demonstrate close agreement within the target (median %$\Delta D_{LOCAL}$: 0.61%, $95^{th}$ percentile range: [-0.20%, 1.34%], as shown in Table 5). This agreement deteriorates with increasing distance from the implant leading to a general dose overestimation by ACE, also evident in Figure 5(b) and Table 5, where % $\Delta D_{LOCAL}$ distributions for structures relatively away from the implant exhibit small (<1.08%) but consistently positive medians and positive skewness, except for the bones. For the bones, ACE exhibits a noticeable dose underestimation relative to the reference MC data with a median % $\Delta D_{LOCAL}$ value of -11.85% ($95^{th}$ percentile range: [-18.33%, -5.14%]). The observed underestimation in bone for test case A is influenced, in part, by the different calcium content of cortical bone used for ACE and MC dose calculations which were 22.5%[20] and 13.2%,[34] respectively. The ray effects and the pattern due to the switch between regions of the multiresolution Cartesian calculation grid used by ACE shown in Figure 5a, as well as the discussed dose overestimation close to the geometry boundaries, and the dose underestimation in bone, are in accordance with findings of initial MBDCA validation studies,[5,21,54] and amendments have been proposed for the latter two.[58,59] These results are consistent with the DVH indices calculated using Oncentra Brachy TPS (Table S6 of the Supplementary Material), where ACE at both HA and SA levels shows a general agreement within 4.95% with MCNP, except for the bones, where it generally underestimates the dose by up to 11.39% and 12.87%, respectively. Larger differences are observed between TG-43 and MCNP for all structures apart from the bones, where the differences remained within 6.98%. Corresponding comparisons between Acuros BV and MCNP for test case A in Figure 5(c) show a general agreement within MC Type A uncertainty apart from the bones, with % $\Delta D_{LOCAL}$ median values and 95% percentile ranges in Figure 5(d) and Table 5 ranging from -0.01% [-1.95%, 0.67%] for the target to 0.53% [-0.65%, 1.84%] for the bowel. For the bones, Acuros BV exhibits a noticeable dose underestimation, with a median % $\Delta D_{LOCAL}$ value of -5.45% (95th percentile range [-11.35%, -0.14%]), while an increased localized dose overestimation is observed just outside the contour of the structure, likely due to volume averaging effects. This underestimation is mainly attributed to the absence of Ca in the cartilage elemental composition used in Acuros BV,[32] whereas MCNP incorporates a 13.2% Ca content in the corresponding material composition,[34] resulting in substantial differences in the mass energy absorption coefficients used by two dose calculation methods. These findings are in accordance with the results of initial MBDCA validation studies.[5,11,21] The corresponding DVH indices calculated using BrachyVision TPS (Table S8 of the Supplementary Material) corroborate these results, showing agreement between Acuros BV and MCNP within 0.57% across all structures, excluding the bones, where Acuros BV underestimates the dose by up to 4.35%.

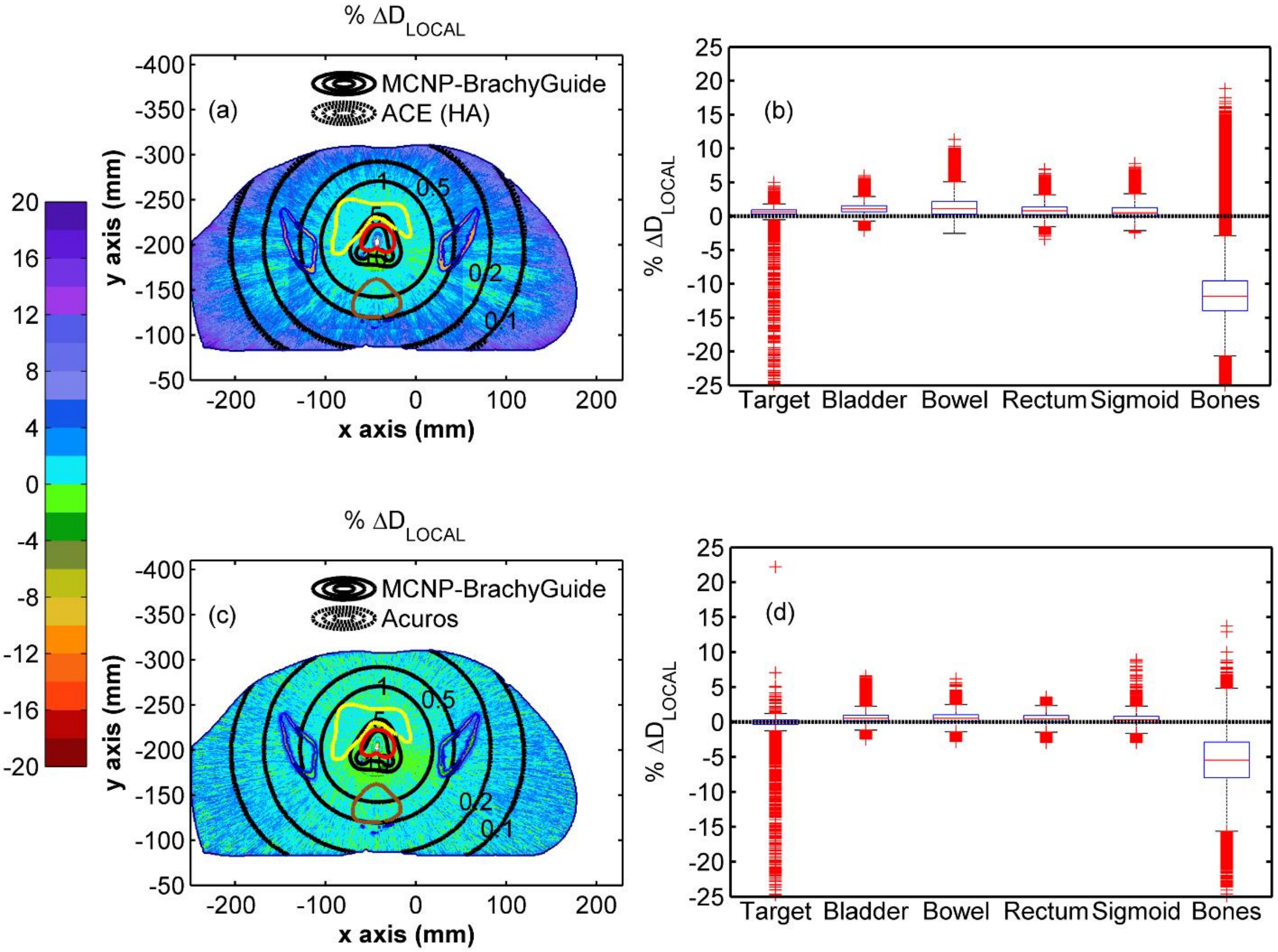

Figure 5. Colormap representations of the % $\Delta D_{LOCAL}$ between (a) ACE (HA) and (c) Acuros BV with MCNP results on an axial slice of Test case A, with selected isodose lines in Gy (0.1, 0.2, 0.5, 1.0, 5.0, 10.0) superimposed (red contour: target, yellow contour: bladder, brown contour: rectum, blue contour: bones, magenta contour: external). Differences in isodoses are not visible due to their overlap. Corresponding box and whiskers plots of the % $\Delta D_{LOCAL}$ between (b) ACE (HA) and (d) Acuros BV with MCNP results are presented for the target, bladder, bowel, rectum, sigmoid, and bones. Whiskers extend to 1.5 times the interquartile range and the red columns shown are formed by the overlapping of outliers.

### 2.5.2 Test case B

Figure 6 presents results from the comparisons between ACE (HA) and reference MCNP data for test case B. Despite its inherent tendency to overestimate dose with increasing distance from the source dwell positions (Figure 6(a)), ACE generally agrees with MCNP within MC Type A uncertainty for the soft tissue-like structures excluding rectum, with median % $\Delta D_{LOCAL}$ values and 95% percentile ranges in Figure 6(c) and Table 5 spanning from 0.36% [-1.32%, 1.79%] for the target to 1.10% [-1.11%, 3.47%] for the bladder. For the rectum, ACE demonstrates a higher dose overestimation compared to MCNP with a median % $\Delta D_{LOCAL}$ value of 1.76% (95th percentile range [-2.50%, 11.44%]). This overestimation primarily arises from ACE modeling the rectum, a heterogeneous structure containing air voxels, as uniformly composed of female soft tissue, resulting in higher mass energy absorption coefficients in those regions relative to MCNP. It should be noted that corresponding DVH indices calculated for the rectum using the Oncentra Brachy TPS (Table S7 of the Supplementary Material) show that this overestimation by ACE remains within 3.87% at both SA and HA levels when compared to MCNP, whereas for TG-43, it exceeds 5% for the majority of indices considered. Although ACE was expected to underestimate dose within the pelvic bones, this effect is less pronounced in test case B compared to test case A due to the smaller bone size, with further suppression resulting from the reduced Ca content used in the MC simulation, reaching up to 15.9%,[34] compared to the 22.5% considered in ACE,[20] which leads to a median % $\Delta D_{LOCAL}$ value in Figure 6(c) and Table 5 of -0.68% (95$^{th}$ percentile range [-11.89%, 9.44%]). For both femoral heads, ACE exhibits a dose underestimation compared to MCNP, with median % $\Delta D_{LOCAL}$ values down to -4.56% (95th percentile range [-10.02%, 3.94%]).

This underestimation reflects the absence of Ca in the mean gland material composition assigned to the femoral heads in ACE, whereas the Ca content in MCNP ranged from 0% to 8.3%. For the bone marrow, which consists of voxels with a median mass density of 1.08 g/cm$^3$ (95th percentile range [0.98, 1.27]) that more closely resembles soft tissue than skeletal tissue, the use of the mean gland material composition in ACE yielded dosimetric results in close agreement with MCNP, with a median % $\Delta D_{LOCAL}$ value of -0.39% (95th percentile range [-9.37%, 5.69%]). These findings align with the DVH results presented in Table S7 of the Supplementary Material, where ACE at both SA and HA levels demonstrates an underestimation within the femoral heads of up to 6.13% and 6.11%, respectively, compared to MCNP, which exceeded the corresponding discrepancies found with TG-43 calculations. This underestimation is generally more pronounced than that observed in the bone marrow and pelvic bones.

Percentage local differences between Acuros BV and reference MC data for test case B are also presented in Figure 6. In Figure 6(b) agreement within MC type A uncertainty can be observed between Acuros BV and MCNP for the soft tissue-like materials apart from the rectum, with median % $\Delta D_{LOCAL}$ values and 95% percentile ranges shown in Figure 6(d) and Table 5 ranging from -0.38% [-3.20%, 1.18%] for the target to 0.81% [-1.97%, 4.39%] for the bowel. Although Acuros BV is expected to assign the elemental composition of air to the air voxels within the rectum,[32] the elevated % $\Delta D_{LOCAL}$ values of approximately 10% observed in these voxels in Figure 6(b) suggest that a soft tissue composition, associated with higher mass energy absorption coefficients than those used in MC simulation, may have been assigned instead. As shown in Figure 6(d) and Table 5, this likely contributes to the overall dose overestimation by Acuros BV compared to MCNP results within the rectum, with a resulting median value of 2.61% (95th percentile range [-1.67%, 12.01%]). These findings reflect the comparison of DVH indices presented in Table S9 of the Supplementary Material calculated using BrachyVision TPS, where Acuros BV shows an overestimation within the rectum of up to 5.77% compared to MCNP. Box plots of the percentage local dosimetric differences shown in Figure 6(d) for the bilateral femoral heads, bone marrow and pelvic bones reveal that combining the material compositions of skeletal muscle and cartilage for the femoral heads and bone marrow, and the cartilage and cortical bone for pelvic bones, yields a dosimetric accuracy for Acuros BV comparable to MCNP with resultant median $\%\Delta D_{LOCAL}$ values of -1.43% (95th percentile range [-5.94%, 4.95%]), 0.83% (95th percentile range [-4.96%, 6.35%]), and –0.14% (95th percentile range [-6.59%, 6.33%]), respectively, in Table 5. These results agree with the corresponding DVH indices summarized in Table S9 of the Supplementary Material, showing that percentage differences between Acuros BV and MCNP for skeletal tissues remain within 3.13%.

Table 5. Comparison of target and critical organs-related dosimetry for test cases A and B performed for MC and TPS MBDCA validation in the form of $\%\Delta D_{LOCAL}$ calculated independently from the TPSs with respect to the reference MCNP results.

| | $\%\Delta D_{LOCAL}$ Median [5$^{th}$ percentile,95$^{th}$ percentile] | | | |
|---|---|---|---|---|
| | MC data validation | | TPS MBDCA data validation | |
| ROI | RapidBrachyMCTPS | eb_gui | ACE (HA) (OncentraBrachy) | Acuros BV (BrachyVision) |
| **Test case A** | | | | |
| Target | -0.06 [-0.92,0.77] | -0.42 [-0.98,0.12] | 0.61 [-0.20,1.34] | -0.01 [-1.95,0.67] |
| Bladder | -0.52 [-2.14,1.11] | -0.47 [-1.47,0.53] | 1.07 [0.00.2.29] | 0.52 [-0.49,1.64] |
| Bowel | -0.57 [-2.91,1.83] | -0.46 [-1.80,0.88] | 1.08 [-0.61,4.30] | 0.53 [-0.65,1.84] |
| Rectum | -0.33 [-2.53,1.85] | -0.49 [-1.73,0.76] | 0.75 [-0.59,2.40] | 0.44 [-0.71,1.62] |
| Sigmoid | -0.56 [-2.50,1.48] | -0.45 [-1.61,0.73] | 0.48 [-0.84,2.84] | 0.33 [-0.88,1.60] |
| Bones | -0.56 [-4.87,3.94] | -0.59 [-2.60,1.41] | -11.85 [-18.33,-5.14] | -5.45 [-11.35,-0.14] |
| **Test case B** | | | | |
| Target | -0.61 [-2.04,0.86] | -0.46 [-1.75,0.89] | 0.36 [-1.32,1.79] | -0.38 [-3.20,1.18] |
| Bladder | -0.52 [-3.05,2.05] | -0.48 [-2.66,1.74] | 1.10 [-1.11,3.47] | 0.71 [-1.44,3.12] |
| Bowel | -0.36 [-3.69,3.23] | -0.57 [-3.38,2.29] | 0.82 [-2.06,5.34] | 0.81 [-1.97,4.39] |
| Rectum | -0.60 [-3.81,2.78] | -0.38 [-3.15,2.47] | 1.76 [-2.50,11.44] | 2.61 [-1.67,12.01] |
| Sigmoid | -0.40 [-3.35,2.80] | -0.47 [-3.02,2.15] | 0.37 [-2.21,3.91] | 0.60 [-2.04,3.72] |
| Right FH | -0.69 [-4.74,3.34] | -0.64 [-3.81,3.07] | -4.20 [-9.58,4.11] | -1.44 [-5.73,4.48] |
| Left FH | -0.49 [-4.62,3.62] | -0.38 [-3.62,3.40] | -4.56 [-10.02,3.94] | -1.43 [-5.94,4.95] |
| Marrow | -0.47 [-4.64,3.83] | -0.45 [-3.67,3.31] | -0.39 [-9.37,5.69] | 0.83 [-4.96,6.35] |
| Pelvic Bones | -0.52 [-4.68,3.78] | -0.48 [-3.73,3.14] | -0.68 [-11.89,9.44] | -0.14 [-6.59,6.33] |

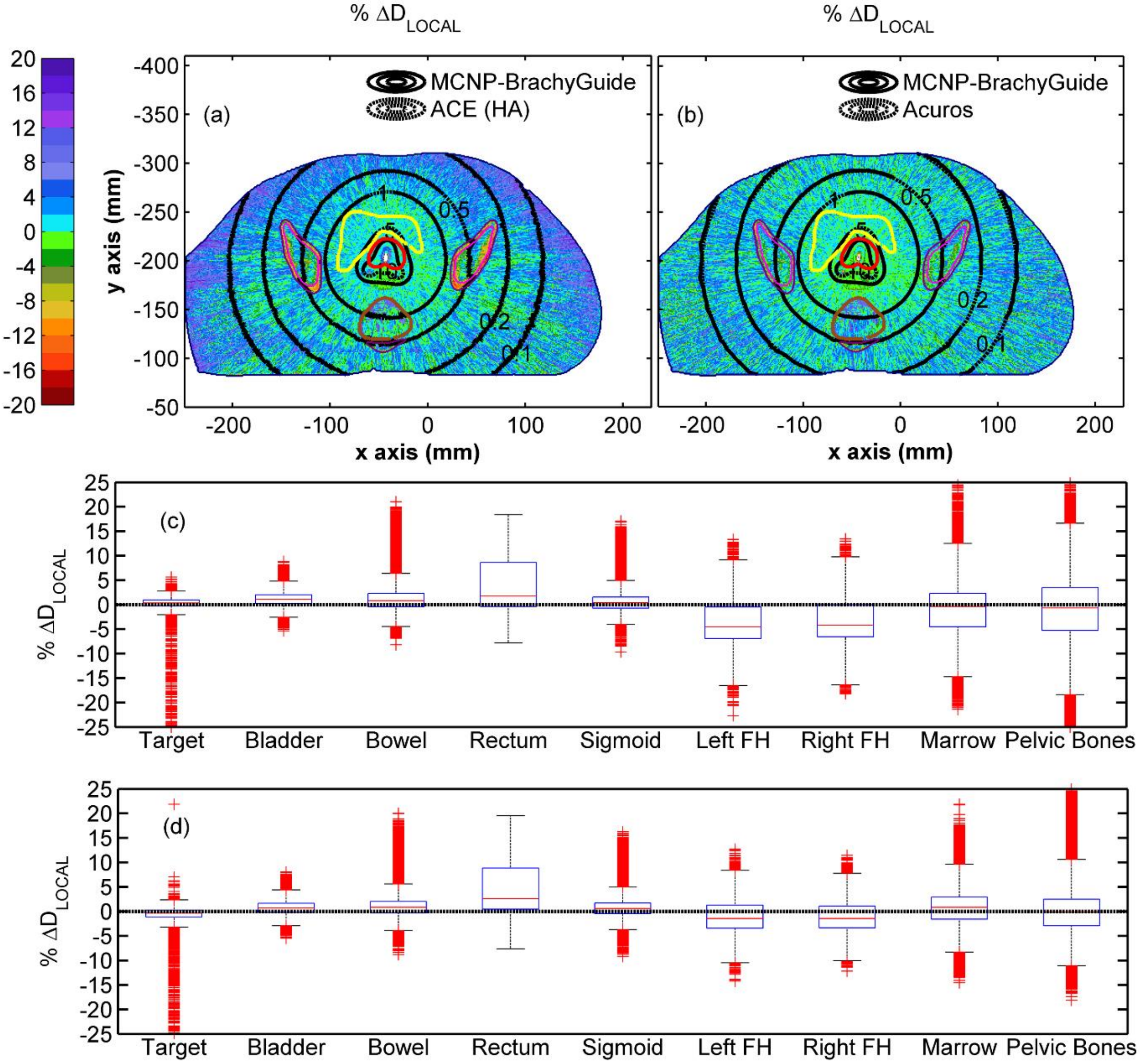

Figure 6. Colormap representations of the % $\Delta D_{LOCAL}$ between (a) ACE (HA) and (b) Acuros BV with MCNP results on an axial slice of Test case B, with selected isodose lines in Gy (0.1, 0.2, 0.5, 1.0, 5.0, 10.0) superimposed (red contour: target, yellow contour: bladder, brown contour: rectum, magenta contour: marrow, blue contour: pelvic bones). Differences in isodoses are not visible due to their overlap. Corresponding box and whiskers plots of the % $\Delta D_{LOCAL}$ between (c) ACE (HA) and (d) Acuros BV with MCNP results are presented for the target, bladder, bowel, rectum, sigmoid, left femoral head, right femoral head, marrow and bones. Whiskers extend to 1.5 times the interquartile range and the red columns shown are formed by the overlapping of outliers.

## 3 DATA FORMAT AND USAGE NOTES

Test cases A and B are hosted on the Zenodo repository (https://doi.org/10.5281/zenodo.15720996), and are accessible also via the Brachytherapy Source Registry, a resource jointly managed by the AAPM and IROC Houston.[60]

For each TPS currently incorporating an MBDCA (Oncentra Brachy by Elekta and BrachyVision by Varian), a set of files is available for download for each test case, including:

1. "Reference TPS" containing 111 CT images of the patient model, the RT structure set (RS), RT plan (RP) and corresponding RT dose (RD) calculated using the MBDCA algorithm, in DICOM RT format.
2. "Reference MC" containing the same files as above but with RD data corresponding to reference results of MC simulations using MCNP. The reference dose distribution is common to both TPSs.
3. User Guides for TPS MBDCA testing using the patient model, in portable document format.
4. "MC input files" containing all the files necessary to perform the MC simulations presented in this work. The MC input files are common to both TPSs.

For Oncentra Brachy, an additional link is provided in the corresponding user guides to download a compressed archive containing XML-formatted data required to configure the generic WG source.

Following the workflow outlined in each TPS-specific user guide for each test case, the end user will be able to perform the commissioning workflow presented in WGDCAB report 372[2] in fulfillment of TG-186 recommendations,[20] based on the analysis of clinically relevant DVH indices that describe high-dose regions and small volumes, as well as intermediate- and low-dose regions involving non-small volumes, consistent with the advanced standard for dose reporting in the ICRU 89 recommendations.[36] This includes the calculation of $DVH_{TPS,User}$, which will be compared with the corresponding $DVH_{TPS,ref}$ and $DVH_{MC,ref}$ by evaluating the percentage local differences $\%\Delta D_{LOCAL}^{TPS}$ and $\%\Delta D_{LOCAL}^{MC}$, respectively. Results of the $DVH_{TPS,ref}$ and $DVH_{MC,ref}$ indices obtained for each test case using the Oncentra Brachy and BrachyVision TPSs along with corresponding comparisons, are provided in Tables S6 – S9 of the Supplementary Material. $DVH_{TPS,ref}$ results for Oncentra Brachy were calculated using ACE in both standard and high accuracy levels, as well as TG-43. For BrachyVision, $DVH_{TPS,ref}$ data were derived using Acuros BV and TG-43. It should be noted that, $DVH_{TPS,ref}$ data distributed with the test cases and included in the corresponding user guides of Oncentra Brachy and BrachyVision TPSs involve MBDCA dose calculations with ACE in high accuracy level and Acuros BV, respectively. Provided that the user's TPS version matches the one used for the reference acquisition, the tolerance for the $\%\Delta D_{LOCAL}^{TPS}$ in Level 1 commissioning should be within 0.1% to ensure that the MBDCA has been properly set up.[2] For Level 2 commissioning, the $\%\Delta D_{LOCAL}^{MC}$ should agree with the corresponding differences between the $DVH_{MC,ref}$ and $DVH_{TPS,ref}$ results within rounding errors, for all calculated indices.[2]

# 4 DISCUSSION

The datasets of this work are the first ever pertinent to the commissioning of MBDCAs for intracavitary $^{192}$Ir HDR treatment of cervical cancer. Similar to the clinically oriented test cases already prepared by the WGDCAB,[24,25] the provided resources and proposed methodology can act as a comprehensive commissioning framework for institutions adopting TPSs that currently utilize MBDCAs. Besides commissioning and periodic QA, the datasets can be employed to assess the benefits of using MBDCAs through comparisons of MBDCA and conventional dose-to-water calculations, intercomparisons of different MBDCAs, and to gain insight into the mechanics of MBDCA implementation given the necessary trade-off between accuracy and speed.

Following previous findings,[24] results of the three MC codes used to validate the reference dose distribution of the datasets agreed within statistical uncertainty with minor exceptions (voxels partially occupied by the source and a limited number of voxels in the periphery of bones in case B, explained in Section 2.4). Hence, the MC results of the datasets can be used for benchmarking purposes by researchers employing MC methods for dosimetry in brachytherapy. Developers of new MBDCAs or MBDCA updated versions can also benefit from these new datasets for one of the most common clinical brachytherapy applications. An example of an area wherein the datasets could be useful for MC researchers or developers is material assignment. While agreement between MBDCA and reference results for soft tissues was influenced only by ray effects in the former and statistical uncertainty in the latter as expected,[19,21] for bony structures substantial differences were observed. These differences increase with distance from the implant as photon energy decreases and the mass energy absorption coefficients of skeletal tissues vary significantly with variable proportion of osseous tissue to bone marrow,[33,34,61] and depend on the number of materials available for assignment as well as the particular elemental compositions assigned. While the necessity of obtaining detailed tissue elemental composition data as input for MBDCAs has been acknowledged and various analytical and patient-specific methods have been suggested,[33] these have not been integrated into clinical practice. The introduction of artificial intelligence-powered software in brachytherapy to assist in OAR delineation, enabling more accurate elemental composition assignment, appears to be a promising alternative for reconciling dose calculation accuracy with clinical time constraints.[62,63] Since the developed datasets include dose data in terms of $D_{m,m}$ which is the proposed dose reporting quantity for both external beam and brachytherapy,[64] they could be extended to include dose data from external beam to serve as a benchmark for methods or tools to accurately assess the cumulative dose in a well-defined, reference geometry.

The patient models and methods developed in this work proved sensitive in identifying the benefits and potential limitations of the commercially available MBDCAs that could affect dosimetric accuracy in gynecological cases treated with intracavitary $^{192}$Ir HDR brachytherapy. The proposed procedure also supported the interpretation of commissioning results within the context of real clinical scenarios, represented by test case B, which appeared less influenced by the identified limitations, yet still underscored the need for strategies to address them. Limitations of the datasets include their specificity in terms of radionuclide, source model, and clinical scenario. Successful commissioning or QA, however, is expected to apply to different sources of the same radionuclide, provided these have been accurately defined in the TPS. The definition of the applicator as a ROI of given material (PPSU) precludes the use of available applicator libraries, thus mitigating potentially introduced bias to the commissioning. At the same time, however, it renders the datasets ineffective for commissioning applicator libraries. Although the datasets were prepared based on a case of intracavitary $^{192}$Ir HDR brachytherapy using a tandem and ring applicator, results can be sensibly expected to hold for applications involving cases where tandem-ovoids or tandem-mold applicators are used, with and without interstitial needles, as well as treatments of endometrial and vaginal cancers. Finally, the datasets are associated with specific TPS versions. While they remain useful for quantifying potential changes incurred by a new version, for commissioning the responsibility remains with the vendors to check and report that TPS-specific datasets are still valid when a new TPS version is released or provide new reference MBDCA data if this is not the case.[2]

## 5 CONCLUSIONS

This work presents the first clinically oriented test case datasets for $^{192}$Ir intracavitary brachytherapy in cervical cancer, comprising a uniformly structured phantom with homogeneous density regions and a realistic anatomical model. The datasets build upon previously developed clinically relevant test cases by the joint AAPM/ESTRO/ABS/ABG WGDCAB to support MBDCA commissioning. They include CT images of each patient model, along with the corresponding structure sets, treatment plans, and reference MBDCA and MC dose distributions in DICOM RT format for two TPSs. TPS- and case-specific user guides are also provided to assist MBDCA users in commissioning their systems, along with the input files used to generate the MC dose distributions.

The provided datasets and proposed methodology offer a robust foundation to guide institutions in commissioning TPSs that utilize MBDCAs. Additionally, the datasets serve as a benchmark for MC simulations conducted by brachytherapy researchers, facilitate intercomparisons of MBDCA performance using the MC reference data as a benchmark, and provide a quality assurance resource for evaluating future TPS software updates. The presented datasets hold promising potential for assessing the benefits and limitations of MBDCAs in gynecological $^{192}$Ir HDR brachytherapy applications, enabling end users to explore strategies that could enhance their reliability in real clinical scenarios.

**Acknowledgements**

The authors would like to acknowledge Prof. Panagiotis Papagiannis for his insightful discussions and valuable advice. The authors also acknowledge support from the Canadian Institutes for Health Research (CIHR) [RMT, funding reference number 437591], Natural Sciences and Engineering Research Council of Canada (NSERC) [RMT, funding reference numbers 2024-05355], Canada Research Chairs (CRC) program (RMT), the Carleton University Research Office. Prarthana

Pasricha is acknowledged for preparing the mass-energy absorption datasets for use in the egs_brachy simulations.

## Conflict of Interest Statement

The authors declare that they do not have any pertinent conflicts of interest.

## Data availability Statement

The test cases developed in this study are available at https://doi.org/10.5281/zenodo.15720996.